\documentclass{aastex631} 

\usepackage{color}

\usepackage{amssymb}

\usepackage{rotating}
\usepackage{mathtools}
\usepackage{comment}
\usepackage{bm}
\usepackage{epstopdf}
\usepackage{amssymb}

\usepackage{rotating}

\usepackage{array}
\newcolumntype{H}{>{\setbox0=\hbox\bgroup}c<{\egroup}@{}}
\usepackage[caption=false]{subfig}

\newcommand{\kms}{\mbox{km~s$^{-1}$}}

\newcommand{\my}{\mbox{$M_{\odot}$~yr$^{-1}$}}
\newcommand{\ls}{\mbox{$L_{\odot}$}}

\newcommand{\ms}{\mbox{$M_{\odot}$}}
\newcommand{\me}{M${_\oplus}$}

\newcommand{\vdensunit}{cm$^{-3}$}

\newcommand{\fbolunit}{\mbox{erg\,s$^{-1}$\,cm$^{-2}$}}

\newcommand{\gsim}{\raisebox{-.4ex}{$\stackrel{>}{\scriptstyle\sim}$}}

\DeclareCaptionFormat{cont}{#1 (cont.)#2#3\par}

\shorttitle{The central star of the Ring Nebula}
\shortauthors{Sahai et al.}

\begin{document}
\title{JWST observations of the Ring Nebula (NGC 6720): III. A dusty disk around its Central Star}

\author{Raghvendra Sahai}
\affiliation{Jet Propulsion Laboratory, California Institute of Technology, Pasadena, CA 91109, USA}
\author{Griet Van de Steene}\affiliation{Royal Observatory of Belgium, Ringlaan 3, B-1180 Brussels, Belgium}
\author{Peter van Hoof}\affiliation{Royal Observatory of Belgium, Ringlaan 3, B-1180 Brussels, Belgium}
\author{Albert Zijlstra}\affiliation{Jodrell Bank Centre for Astrophysics, Department of Physics \&\ Astronomy, The University of Manchester, Oxford Road, \\
Manchester M13 9PL, UK}
\author{Kevin Volk}\affiliation{Space Telescope Science Institute, 3700 San Martin Drive, Baltimore, MD 21218, USA}
\author{Harriet L. Dinerstein}\affiliation{University of Texas at Austin, Austin, TX 78712, USA}
\author{Michael J. Barlow}\affiliation{Department of Physics and Astronomy, University College London, Gower Street, London WC1E 6BT, UK}
\author{Els Peeters}\affiliation{Department of Physics and Astronomy, University of Western Ontario, London, Ontario, Canada}\affiliation{Institute for Earth and Space Exploration, University of Western Ontario, London, Ontario, Canada}\affiliation{SETI Institute, Mountain View, CA, USA}
\author{Arturo Manchado}\affiliation{Instituto de Astrof\'{\i}sica de Canarias, 38200 La Laguna, Tenerife, Spain}\affiliation{Universidad de La Laguna (ULL), Astrophysics Department, 38206 La Laguna, Tenerife, Spain}\affiliation{CSIC, Spain}
\author{Mikako Matsuura}\affiliation{Cardiff Hub for Astrophysics Research and Technology (CHART), School of Physics and Astronomy, Cardiff University,
The Parade, Cardiff CF24 3AA, UK}
\author{Jan Cami}\affiliation{Department of Physics and Astronomy, University of Western Ontario, London, Ontario, Canada}\affiliation{Institute for Earth and Space Exploration, University of Western Ontario, London, Ontario, Canada}\affiliation{SETI Institute, Mountain View, CA, USA}
\author{Nick L. J. Cox}\affiliation{ACRI-ST, Centre d’Etudes et de Recherche de Grasse (CERGA), 10 Av. Nicolas Copernic, 06130 Grasse, France}\affiliation{INCLASS Common Laboratory., 10 Av. Nicolas Copernic, 06130 Grasse, France}
\author{Isabel Aleman}\affiliation{Laborat\'{o}orio Nacional de Astrof\'{i}sica, Rua dos Estados Unidos, 154, Bairro das Na\c{c}\~{o}es, Itajub\'{a}, MG, CEP 37504-365, Brazil}
\author{Jeronimo Bernard-Salas}\affiliation{ACRI-ST, Centre d’Etudes et de Recherche de Grasse (CERGA), 10 Av. Nicolas Copernic, 06130 Grasse, France}\affiliation{INCLASS Common Laboratory., 10 Av. Nicolas Copernic, 06130 Grasse, France}
\author{Nicholas Clark}\affiliation{Department of Physics and Astronomy, University of Western Ontario, London, Ontario, Canada}
\author{Kay Justtanont}\affiliation{Chalmers University of Technology, Department of Space, Earth and Environment, Onsala Space Observatory, S-439 92 Onsala, Sweden}
\author{Kyle F. Kaplan}\affiliation{University of Texas at Austin, Austin, TX 78712, USA}
\author{Patrick J. Kavanagh}\affiliation{Department of Experimental Physics, Maynooth University, Maynooth, Co Kildare, Ireland}
\author{Roger Wesson}\affiliation{Department of Physics and Astronomy, University College London, Gower Street, London WC1E 6BT, UK}\affiliation{Cardiff Hub for Astrophysics Research and Technology (CHART), School of Physics and Astronomy, Cardiff University, The Parade, Cardiff CF24 3AA, UK}

\begin{abstract}
The planetary nebula NGC\,6720, also known as the ``Ring Nebula", is one of the most iconic examples of nearby planetary nebulae whose morphologies present a challenge to our theoretical understanding of the processes that govern the deaths of most stars in the Universe that evolve on a Hubble time. We present new imaging with JWST of the central star of this planetary nebula (CSPN) and its close vicinity, in the near- to mid-IR wavelength range. We find the presence of a dust cloud around the CSPN, both from the spectral energy distribution at wavelengths $\gsim$5\,\micron, as well as radially-extended emission in the 7.7, 10 and 11.3\,\micron~images. From modeling of these data, we infer that the CSPN has a luminosity of 310\,\ls, and is surrounded by a dust cloud with a size of $\sim$2600\,au, consisting of relatively small amorphous silicate dust grains (radius $\sim$0.01\,\micron) with a total mass of $1.9\times10^{-6}$\,\me. However, our best-fit model shows a significant lack of extended emission at 7.7\,\micron~-- we show that such emission can arise from a smaller ($7.3\times10^{-7}$\,\me) but uncertain mass of (stochastically-heated) ionized PAHs. However, the same energetic radiation also rapidly destroys PAH molecules, suggesting that these are most likely being continuously replenished, via the outgassing of cometary bodies and/or the collisional grinding of planetesimals. We also find significant photometric variability of the central source that could be due to the presence of a close dwarf companion of mass $\le$0.1\,\ms.

\end{abstract}


\section{Introduction}\label{intro}
The presence and origin of dusty disks around main-sequence (MS) and pre-MS stars is well understood. These are first seen as the gas and dust-rich planet-forming disks in young stellar objects (YSOs) and are an integral part of the star-formation process itself, and then as the gas-poor debris disks around MS stars resulting from the collisions of large planetesimals that produce second-generation dust particles \citep{Rieke05}. The dust in these debris disks dissipates long before the stars evolve off the MS.

Remarkably, dusty disks or disk-like structures manifest themselves again as these stars reach the ends of their lives as Asymptotic Giant Branch (AGB) stars, post-AGB stars and the central stars of planetary nebulae (PNe) (e.g., \citealt{Sahai2007, Sahai2011, Hillen2017}). AGB stars, representing the death throes of stars with MS masses of $\sim1-8$\,\ms, are very luminous ($L\sim5000-10,000$\,\ls) and cool ($T_{eff} < 3000$\,K) and experience heavy mass-loss (with rates up to $10^{-4}$\,\my, see e.g., review by \citealt{Decin2021}) that depletes most of the stellar envelope and accelerates their evolution to the PN phase, through a transitory post-AGB phase. These stars evolve to higher temperatures through the post-AGB and PN phases at almost constant luminosity, fading and becoming white dwarfs (WDs) at the ends of their lives. It is during these post-AGB and WD phases, that the presence of disk-like structures around the central stars becomes observationally apparent once again, raising questions about their nature, formation, longevity and potentially a second phase of planet-formation. The disks have a large range of sizes, found from direct imaging or derived from modeling the spectral energy distribution (SED). These disks range from very small disks found around cool central WD stars ($\lesssim0.01$\,AU) (e.g., \citealt{Ballering2022}) to much larger disks extending to radii up to $\sim1000$\,AU that have been found in AGB stars (e.g., \citealt{Sahai2022}), post-AGB stars (e.g., \citealt{DeRuyter06}), and the central stars of planetary nebulae (e.g., \citealt{Su2007, Chu11, Bilikova12, Sahai2023}).

We report here new JWST observations of the central star (and its immediate environment) of the iconic PN, the Ring Nebula (NGC\,6720), resulting in the discovery of the second resolved dusty disk around a CSPN\footnote{the first being the dusty disk around the CSPN of the Southern Ring, NGC\,3132 \citep{Sahai2023}}. We will hereafter refer to the central star (i.e., the WD) as the CSPN; the CSPN and its immediate environment will be referred to as the central source or CS. This PN, of long-standing interest for both amateur and professional astronomy, has been extensively studied, both from ground-based (e.g., \citealt{balick1992,bryce1994,guerrero1997}) and space-based observatories (\citealt{sahai2012,odell2013}), yet it continues to be an amazing astrophysical laboratory yielding new and unexpected insights into the extraordinary deaths of intermediate-mass stars (e.g., \citealt{Wesson2024,kastner2025}). NGC\,6720 was imaged through a wide suite of filters from 1.6 to 25\,\micron~using the NIRCAM and MIRI instruments on JWST via program ID GO-01558 \citep{Wesson2024}. A log of the observations is provided in Table 1 of \cite{Wesson2024}, who carried out an imaging study of rings, globules and arcs in the nebula. \cite{Wesson2024} found that NGC\,6720's CSPN has two companions. One of these is a distant mid-K spectral type dwarf companion, CSPN(B), based on its having the same parallax and proper motion as NGC\,6720's CSPN, CSPN(A) (\citealt{GonzalezSantamaria21}) with a projected separation of $\sim15,000$\,AU. In addition, there is another possible companion, CSPN(C), that is much closer, with a period of about $280\pm70$ yr, and thus at a separation of $50\pm15$\,AU, inferred from the presence of low-contrast, regularly spaced concentric arc-like features seen in the F770W, F100W and F1130W images of the nebula. The \cite{Wesson2024} study is the first publication of several studies of this object using data from JWST program GO-01558 that include this paper, and two studies of key diagnostic regions of the nebula, one focussing on the PAH emission (Clark et al. 2024, submitted), and another on the rich H$_2$ emission-line spectrum (van Hoof et al. in preparation).

The paper is organised as follows. In \S\,\ref{obs_desc} we describe the imaging observations of the CS of NGC\,6720, in \S\,\ref{making-sed} we describe the construction of the full SED of the CS from UV-to-infrared wavelengths as well as the characterization of the extended mid-IR emission, in \S\,\ref{cspn} we derive the properties of the CSPN from fitting the UV to near-IR SED, in \S\,\ref{models} we model the dust emission, in \S\,\ref{cs-var} we discuss the optical photometric variability of the CS, in \S\,\ref{discuss} we discuss the implications of our results for the origin and formation of the NGC\,6720 CS disk, including the possible presence and role of unseen bound companions, and in \S\,\ref{conclude} we summarize the main conclusions of our study.

\section{The Central Star of NGC\,6720 and its Near Environment}\label{obs_desc}
The extended morphology of the nebula from 2 to 21\,\micron~shows that the CS is located within a roughly circular region of radius about $25{''}$
that is of relatively low-surface brightness in the NIRCAM images, as well as the MIRI images in most of the filters (F560W, F770W, F1130W, F1280W, F1500W, and F2100W) (Fig.\,\ref{n6720-csregion}, see also Figs.\,1, 4 \& 5 of \citealt{Wesson2024}). The exceptions are the F1000W, F1800W, and F2550W filters, in which the central region is almost filled, and almost as bright (F1800W) or brighter than the main ring (F1000W, F2550W), except for a roughly linear structure that is relatively ``dark" and lies approximately along the
major axis of the nebula {(Fig.\,\ref{n6720-csregion}). The F1000W filter includes strong gas emission lines of [SIV] and [ArIII] in its bandpass; the F1800W includes a strong contribution from [SIII]. The F2550W filter includes the strong [O IV] line. The CS (Fig.\,\ref{n6720-cs-hst-jwst}) is located within the linear structure, and is generally well-isolated from surrounding nebulosity in filters with central wavelengths shortwards and including 7.7\,\micron, but lies on the edge of bright nebular\footnote{we use the term ``nebular" here and elsewhere to mean ``belonging to larger structures that are part of the large planetary nebula, and not localised around the CSPN"} emission in the images at longer wavelengths. No localised emission on the CSPN can be seen in filters with central wavelengths longwards of 12.8\,\micron.

\begin{figure}[hbtp]
\includegraphics[width=0.99\textwidth]{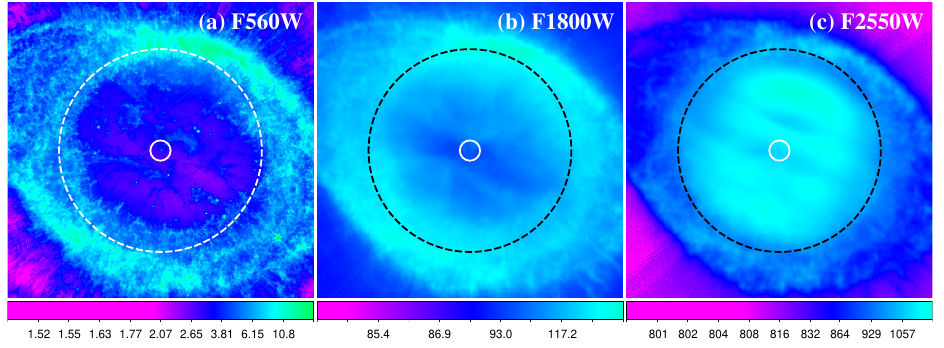}
\caption{JWST/MIRI images showing the extended nebular emission around NGC\,6720's CS in 3 filters: (a) F560W, (b) F1800W, and (c) F2550W.
Large dashed circles (of radius $25{''}$) in each panel show an extended nebular region around the CS (located at center of small circle, or radius $2\farcs5$). Intensity units  (MJy/sr) are shown in the scale bars at the bottom of each image.
}
\label{n6720-csregion}
\end{figure}

\subsection{Spectral Energy Distribution \& Radial Intensities}\label{making-sed}

We have constructed the spectral energy distribution (SED) of the CS over the UV to mid-IR ($\sim 0.09-21$\,\micron) range as follows. The SED in the optical--UV--near-infrared range was determined using archival UV spectra from IUE \footnote{https://archive.stsci.edu/missions/iue} and published optical--near-IR photometry (Table\,\ref{tbl-flux-cs}). For the near- to mid-IR region, we used the JWST imaging data from NIRCAM and MIRI. We extracted photometry of the CS from the NIRCAM images obtained with filters F162M, F212N, F300M and F335M,  using relatively small circular apertures for the CS and annular apertures for the sky background (Table\,\ref{tbl-flux-cs}), with aperture corrections determined using field stars (Table\,\ref{tbl-fs-psf}), as described by \cite{Sahai2023} for the CS of the PN NGC\,3132.

\begin{figure}[hbtp]
\includegraphics[width=0.99\textwidth]{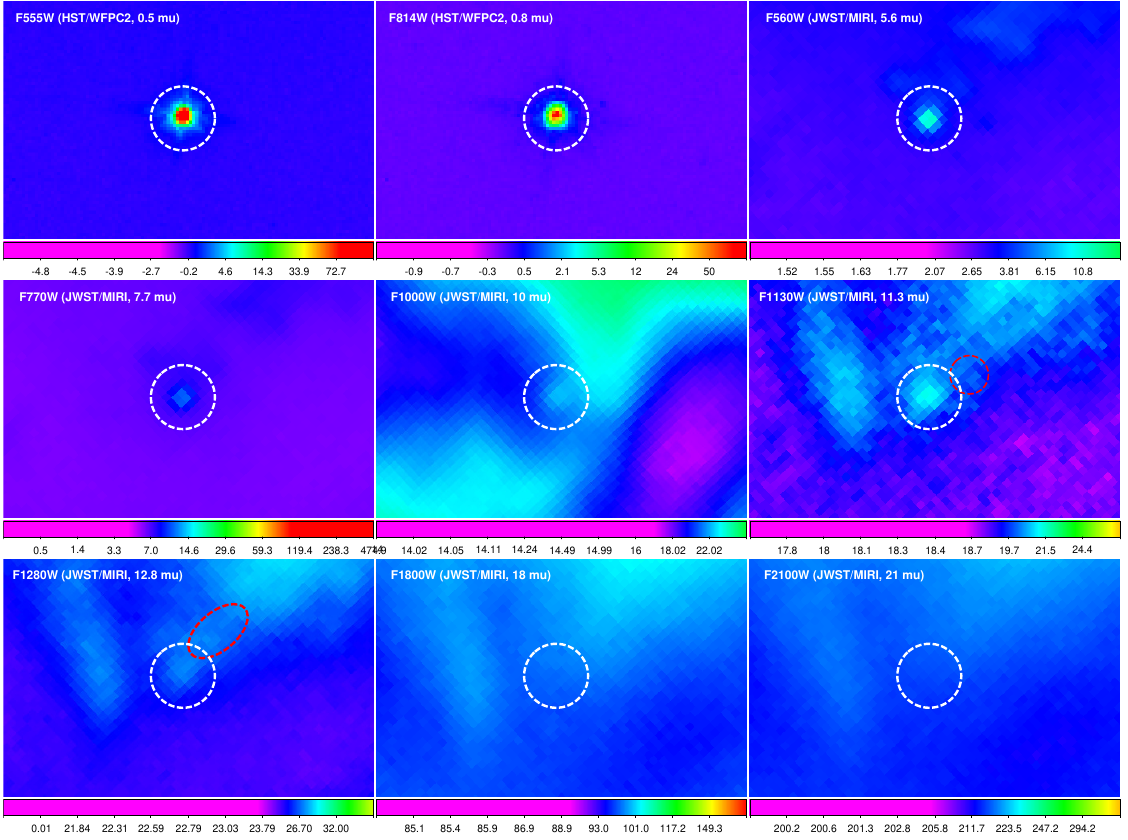}
\caption{Comparison of JWST/MIRI images of the central region of NGC\,6720, taken in 7 filters covering the $5.6-21$\,\micron~wavelength range (F560W, F770W, F1000W, F1130W, F1280W, F1800W, and F2100W), with HST/\,WFPC2 images at optical wavelengths (0.55\,\micron: F555W and 0.81\,\micron: F814W).
White dashed circles (of diameter $1{''}$) locate the central WD star in the images. Intensity units are shown in the scale bars at the bottom of each image: counts/s (cps) per pixel for the HST images (1 cps per pixel is 8.32 MJy/sr and 12.9 MJy/sr, respectively, in the F555W and F814W images) and MJy/sr for the JWST images. Red ellipse in the F1280W image shows the location of an elongated nebular feature, spur(nw), that lies close to and/or overlaps the CS. Red circle in the F1130W image shows a region of nebular emission in the near-vicinity of the CS used for comparing the mid-IR colors of nebular emission with that of the CS.
}
\label{n6720-cs-hst-jwst}
\end{figure}


A different strategy was adopted for the MIRI images, which show the presence of underlying, and/or nebular structures in the near-vicinity of the CS. Images with filters F560W, F770W, F1000W and F1130W show a clear local brightness peak centered on the CS. For these, we (i) subsampled each image by a factor of 3, (ii) extracted a radial intensity distribution, I(r), for each filter by averaging the intensity over an angular wedge with its vertex centered on the CS and a specific angular range chosen to avoid the nebular contamination in the vicinity.

\begin{figure}[hbtp]

\includegraphics[width=0.9\textwidth]{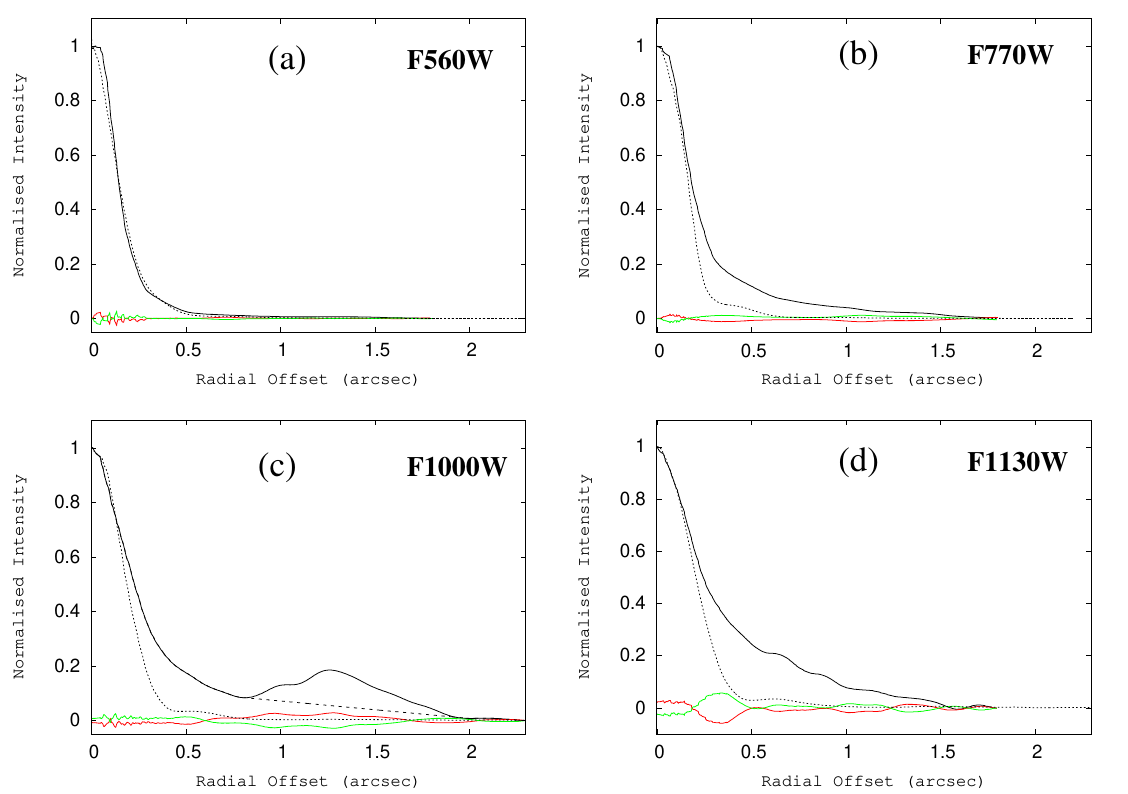}
\caption{Radial intensity distributions extracted from MIRI images in 4 different filters (solid black curves), by averaging the intensity over an angular wedge (full-wedge intensity) with its vertex centered on the CS, with the specific angular range chosen to avoid the nebular contamination in the vicinity in each image. Each  intensity distribution has been normalised by the corresponding peak intensity, which is 7.44, 4.15, 3.06, and 1.8 MJy/sr respectively, for the F560W, F770W, F1000W, and F1130W filters. For each filter, the dotted curves show the PSF extracted from a field star within the field-of-view, and the red and green curves show the differences between the full-wedge intensity and two half-wedge intensities (see text for definition of half-wedge intensity). The broad bump in the  F1000W radial intensity at a radial offset of about $1\farcs25$ is due to a faint nebular structure in the F1000W image that cuts across the angular wedge used for extracting the radial intensity; dashed line shows a linear interpolation of the intensity across the edges of this bump.}
\label{n6720-rcut-miri}
\end{figure}

In all filters shortwards of, and including F560W, the CS appears to be point-like (Figs.\,\ref{n6720-cs-hst-jwst}, \ref{n6720-rcut-miri}). The radial intensity cuts for the F770W, F1000W and F1130W images show a central source with a FWHM comparable to (or slightly larger than) that of the corresponding PSF, together with a weaker ``skirt" of emission at larger radii (Fig.\,\ref{n6720-rcut-miri}). These cuts show that the F770W image has the weakest and least extended nebular emission near the CS, whereas the F1000W image has the strongest and most extended nebular emission near the CS. The PSF for each filter was determined from the corresponding field stars used for the aperture correction for that filter (Table\,\ref{tbl-fs-psf}). The F1000W image has a faint nebular structure that cuts across the selected angular wedge at a radial offset of about $1\farcs25$, resulting in a broad bump in the radial intensity. Hence, for this filter we made a linear interpolation of the intensity across the edges of this bump (dashed line in Fig.\,\ref{n6720-rcut-miri}c). For each of the filters F560W, F770W, F1000W and F1130W, we measure a (i) core flux, $F_{core}$ ($F_{tot}$), by integrating the radial intensity curve to a radius equal to $0.5\times$FWHM of the image PSF in that filter, and (ii) a total flux, by integrating the radial intensity curve to a radius of $1\farcs8$, where the radial intensity is zero. The total flux is significantly lower than the total flux in all filters except F560W, showing that the bulk of the emission at wavelengths $\sim 7.7$\,\micron~and longer comes from the extended component in the CS.

For each filter, in order to assess systematic uncertainties in the radial intensity (the ``full-wedge" intensity), we also extracted two ``half-wedge" intensities, which are the average intensities averaged over two equal contiguous halves that together span the full angular range for that filter. The differences between the full-wedge and half-wedge intensities (red and green curves in Fig.\,\ref{n6720-rcut-miri}) show that these arise at relatively large radial offsets where the emission is relatively low, and therefore much more affected by uncertainties in the  sky background level. The exception to this is the F1130W filter, which also show a local bump centred at $0\farcs37$ in the difference images, with a peak that is $\sim20$\,\% of the full-wedge intensity at that radius.

For the F1280W image, although there is a bright region at the location of the CS, it has roughly the same intensity as, and thus cannot be distinguished from, the nebular spur ``spur(nw)" (marked by a magenta ellipse in F1280W image in Fig.\,\ref{n6720-cs-hst-jwst}). For F1500W, F1800W, and F2550W images, no compact source can be seen at the location of the CS. For these filters (i.e., F1280W--F2550W), we determined upper limits to the fluxes as follows: (i) for each image, we estimated the $1\sigma$ noise in a circular aperture centered at the location of the CS, with diameter equal to the FWHM of the PSF in each filter (as determined from field stars), (ii) assuming that a detectable CS source would have a Gaussian shape with the PSF FWHM and a half-power intensity of 3$\sigma$ in order to be detectable, we computed the flux of this source, and set it to the upper limit on $F_{core}$. The errors in the fluxes are conservative and mostly arise from uncertainties in the sky background.
We also extracted photometry in the images for filters F560W--F1280W from a circular patch in the near-vicinity of the CS (Table\,\ref{tbl-flux-patch}). The colors of this patch are very different from those of the CS.

\section{Stellar Effective Temperature and Luminosity}\label{cspn}
Published values for the effective temoperature of the CSPN vary over the range $T_{eff}\sim(101-162)$ kK (e.g., \citealt{Guerrero2013, Kaler1989}) -- the exact value adopted affects the estimated bolometric flux, and thus the luminosity of the central star. We have adopted an average value of 135\,kK for the WD's effective temperature, $T_{eff}$ in order to fit a model WD spectrum to the SED of the CS in the UV-optical wavelength range (Fig.\,\ref{sed-obs-mod-uv}) as follows.

\begin{figure}[hbtp]

\includegraphics[width=0.6\textwidth]{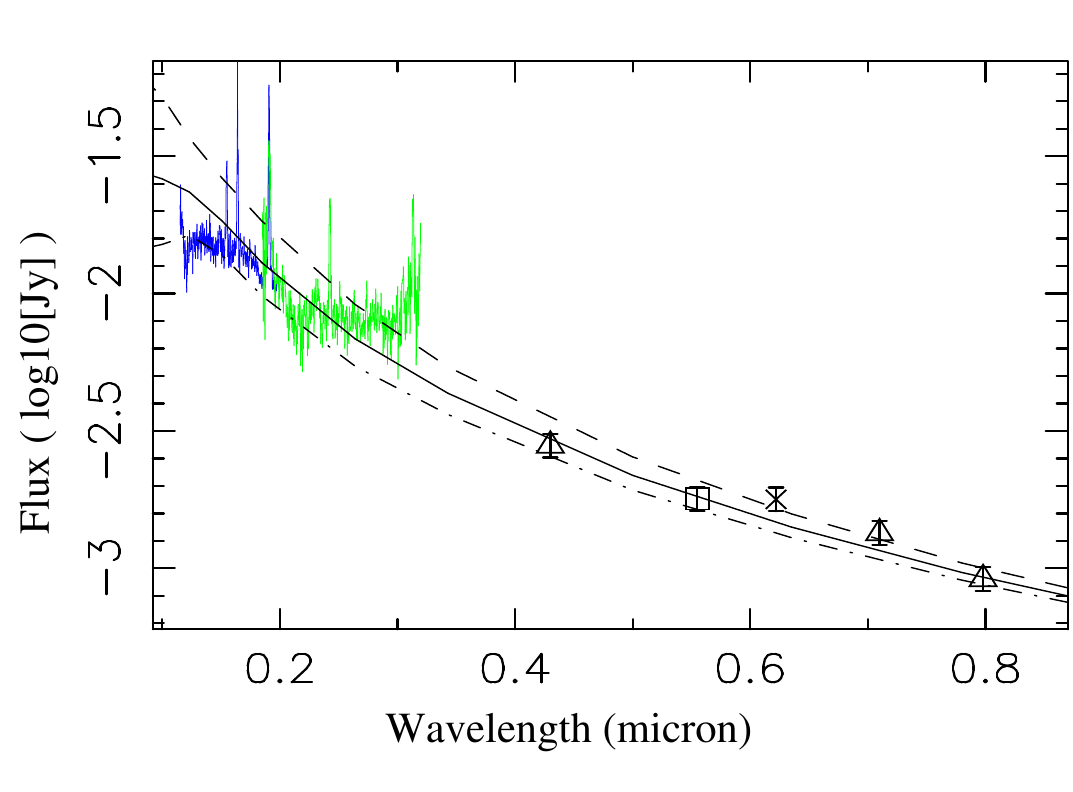}

\caption{Observed UV spectra and optical photometry (black symbols) of the CS of NGC\,6720, together with model SEDs. The model SEDs are that of a white dwarf with $T_{eff}=135,000$\,K, log$g$\,(cm\,s$^{-2})=8$, and luminosity, $L=310$\,\ls, with three different values of the foreground extinction: $A_V=0.15^{\rm mag}$ (solid curve), $A_V=0.0$ (dashed), $A_V=0.27^{\rm mag}$ (dash-dotted). The UV spectra are taken from the IUE MAST archive (data id: SWP\,07230, blue curve, and LWR\,06238, green curve). Error bars on the observed photometry are conservative estimates.}
\label{sed-obs-mod-uv}
\end{figure}

We computed the stellar spectrum using the T\"ubingen NLTE Model Atmosphere Package (TMAP) (\citealt{Rauch2003,Werner2003,Werner2012}) for $T_{eff}=135,000$\,K using solar abundances and log$g$\,(cm\,s$^{-2})=8$. We note that (i) an H-only WD model produces a poorer fit for the fluxes at 3 and 3.35\,\micron, with the model values being significantly higher than the observed fluxes, and (ii) within the wavelength range over which data are available, the WD models are not sensitive to the exact log$g$ value. The interstellar extinction towards NGC\,6720 has been estimated to be $A_V=0.27^{\rm mag}$, based on the value of $c(H\beta)=0.13\pm0.04$ by \cite{odell2009}. However, with $A_V=0.27^{\rm mag}$, the model SED that fits the observed optical photometry, is significantly below the observed IUE spectrum in the UV region (Fig.\,\ref{sed-obs-mod-uv}). We find that $A_V=0.15^{\rm mag}$ produces a much better fit (Fig.\,\ref{sed-obs-mod-uv}) to the FUV -- we have adopted this value, the resulting bolometric flux, $F_{bol}=1.6\times10^{-8}$\,erg\,cm$^{-2}$\,s$^{-1}$ and the luminosity, $L=310$\,\ls, for a distance D=790\,pc based on its trigonometric parallax\footnote{from Gaia DR3 \citep{Gaia2022}, see \cite{Wesson2024}}. We find that the observed SED shows an increasing excess over the observed flux at wavelengths greater than $\sim5$\,\micron. The most likely explanation for this excess is that it arises from the presence of warm/hot dust around the CSPN.

\begin{figure}[hbtp]

\includegraphics[width=0.9\textwidth]{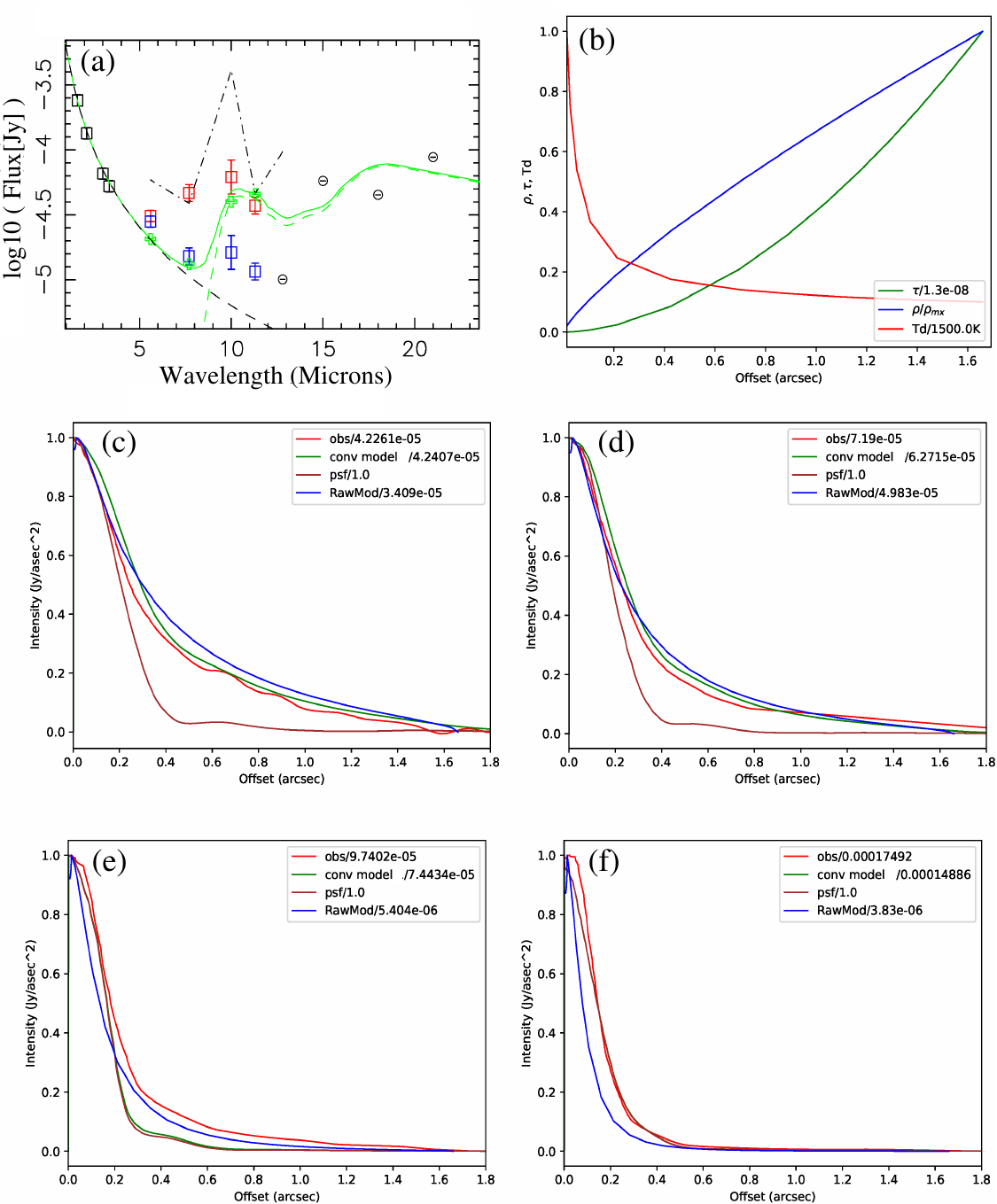}    

\caption{(a) Observed mid-IR photometry (black symbols) and model SED (smooth green curve) for the CS of NGC\,6720 and a circumstellar dust shell with silicate grains, with dust temperature at inner radius of shell $T_{\mathrm d}=1500$\,K, density power-law $\rho(r)\propto r^{0.8}$, outer-to-inner radius ratio $Y=125$ and dust grain-radius 0.01\,\micron. For $\lambda<5$\,\micron, black boxes show the total flux. For $\lambda>5$\micron, red (blue) boxes show the total (core) flux. Circles show upper limits for the core flux. Error bars on the observed photometry are conservative estimates. The dashed green curve shows the thermal emission from dust, and the dashed black curve shows the attenuated starlight. Dash-dot curve shows the relative fluxes of a patch covering a region of nebular emission in the near-vicinity of the CS. Green square symbols show the band-averaged total model fluxes for specific filters. (b) Normalized density, tangential optical depth, and temperature of the dust. Remaining panels show the normalised observed and model (monochromatic, at the center wavelength of each filter) radial intensity distributions of the dust shell for the (c) F1130W, (d) F1000W, (e) F770W, and (f) F560W filters. The numerical division factors in the legends for panels $b-f$ show the values used for normalization.}
\label{sed-obs-mod-sil}
\end{figure}

\begin{figure}[hbtp]
\includegraphics[width=0.9\textwidth]{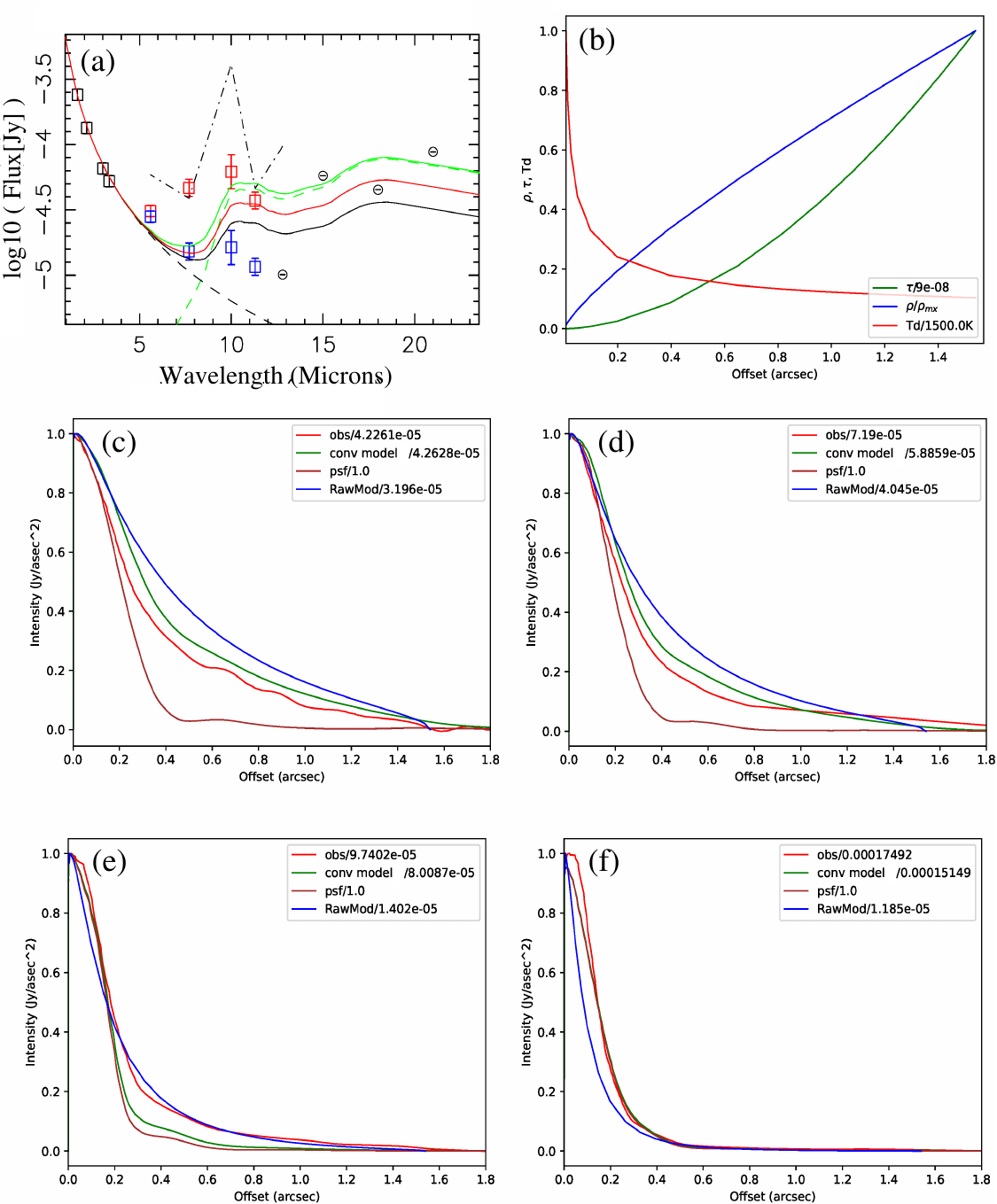}    

\caption{As in Fig.\,\ref{sed-obs-mod-sil}, but with a dust composition that is 50\% silicate and 50\% amorphous carbon.}
\label{sed-obs-mod-silamC}
\end{figure}

\section{Radiative Transfer Modelling of SED and Extended Emission of the CS}\label{models}
We have used the DUSTY and CLOUDY radiative transfer codes \citep{ivezic2012,ferland2017} to model the SED, as well as the radial intensity distributions at 5.6, 7.7, 10 and 11.3\,\micron, of the CS. We need both these codes because (as we show below) we need both thermally and stochastically-heated (i.e., PAHs) dust grains, but the DUSTY code can only model emission from dust grains in thermal equilibrium; the CLOUDY code is needed for modeling the emission from stochastically-heated dust grains. Although it is likely that the dust cloud is disk-like, since we have no direct information about its morphology, we have chosen to use 1D modeling. This does not affect our results because (as we show below) the optical depth of this cloud is $<<1$, even at relatively short wavelengths. Even if the dust cloud had a disk conﬁguration, the radial optical depth near and in the equatorial plane would remain well below unity. Although in principle one should use band-averaged model flux densities for comparsion with the observed photometry, we find that the former are not significantly different from the monochromatic flux densities in wavelength regions where the model spectrum shows a monotonic smooth variation, i.e., for $\lambda\lesssim4$\,\micron. However the model spectra (discussed below) do show strong, non-monotonic variations in the wavelength regions covered by the bandpassses of the the four MIRI filters F560W, F700W, F1000W, and F1130W filters -- hence for these we have used band-averaged model flux densities for comparison with the observations. The input parameters and output properties for our best-fit models are given in Table\,\ref{tbl-dusty}.

\subsection{Thermal Emission: DUSTY}\label{md-dusty}
We first use DUSTY modeling in order to explore the relevant input parameter space because it provides both the SED as well as the radial intensity distributions directly as part of its output. The main input parameters of the DUSTY model are (i) the dust temperature at the inner shell radius ($T_{\mathrm d}$), (ii) the total radial optical depth at 0.55$\micron$~($\tau _V$), (iii) the shell density distribution (iv) the grain-size distribution for a choice of grain composition, (v) the relative shell thickness ($Y$ = ratio of the shell's outer radius,  $R_{ou}$, to its inner radius, $R_{in}$), (vi) the spectrum of the central star -- for this, we use the stellar spectrum that was used to fit the UV-optical-NIR data as described in \S\,\ref{cspn}.

The shell density distribution was assumed to be a power-law, $\rho_d(r)\propto r^{-p}$. For the grain-sizes, we used grains with a fixed radius, $a$, because we found that we had to vary the grain radius in order to find the best-fit. Using, for example, a distribution function for grain radius, such as the Mathis, Rumpl, Nordsieck (MRN) one with $n(a) \propto a^{-q}$ for $a_{\mathrm{min}}$ $\leq a \leq$ $a_{\mathrm{max}}$ (\cite{mathis1977}), would require adjusting the values of 3 different parameters, which would be significantly more poorly constrained, given our fairly limited observational constraints.

Our best-fitting model (Fig.\,\ref{sed-obs-mod-sil}) requires silicate grains in order to fit the shape of the SED in the $5-11.3$\,\micron~region, specifically the local peak at $\sim10$\,\micron. Amorphous carbon grains produce a monotonically-varying, smooth shape that does not fit these data. The dust temperature at the inner shell radius needs to be relatively high, $T_{\mathrm d}\gtrsim1200$\,K, in order to produce a bright core with a width that does not exceed the observed one as seen in the radial intensity distributions; the corresponding value of the inner shell radius is 10.5\,au. The dust density power-law exponent, $p$, is constrained by the extended emission in the radial intensity distributions; we find that $p\sim-0.8$ (i.e., with density increasing outwards with radius) is needed to fit the extended emission seen in the radial intensity distributions for F1000W and F1130W.  We also require the grains to be relatively small, with $a_{\mathrm{max}}\lesssim0.01$\,\micron, in order for these to be warm enough in the extended parts of the dust cloud to produce adequate emission there -- using larger grains results in grain temperatures too low to produce the extended emission. The outer radius of the shell, corresponding to our best-fit model value of $Y=125$, is $\sim1300$\,au.

However, this model shows a significant lack of extended emission for F770W, and the total model flux at 7.7\,\micron~is much less than observed. The model also has inadequate emission to fit the F560W photometry well, although the discrepancy is much less than for F770W. We investigated models with mixtures of silicate and amorphous carbon grains, but the resulting best-fit model was worse than the silicate-only model -- specifically, the model radial intensities in the F1130W and F1000W filters provide a significantly worse fit (Fig.\,\ref{sed-obs-mod-silamC}), compared to the silicate-only model. We show in the next section (\S\,\ref{pah-cloudy}), a path forward to help resolve these discrepancies by adding very small grains that can be heated stochastically to much higher temperatures than possible for larger grains in thermal equilibrium.

There are additional small discrepancies between the observed and model fluxes at 10 and 11.3\,\micron~-- the model flux at 10\,\micron~(11.3\,\micron) is slightly below (above) the observed lower (upper) limits on the observed fluxes. A plausible explanation for these discrepancies is that the intrinsic shape of the 10\,\micron~emission feature in the CS is different from the assumed model one; this explanation is supported by the varied shapes of this feature as observed in the dust emission from a sample of WDs \citep{farihi25}. JWST/\,MIRI spectroscopy of the very central regions is needed in order to accurately characterize the shape of the SED and make further progress on understanding the dust cloud around the central star of NGC\,6720.

We find the mass of the dust shell (since DUSTY does not provide a direct measure of the shell dust mass), using Eqn. 1 of \cite{Sahai2023}, i.e.,
\begin{equation} M_{d} = 4\pi\,[(n-1)/(3-n)]\,y(Y)\,R^{2}_{in} (\tau_{10}/\kappa_{10})
\label{eqn-massd}
\end{equation}
where $y(Y)=(Y^{3-n}-1)/(1-Y^{1-n})$, and $\tau_{10}$ and $\kappa_{10}$ are, respectively, the optical depth and the dust mass absorption coefficient at 10\,\micron. For our best-fit model, $\tau _V=1.3\times10^{-8}$ and $\tau_{10}=8.3\times10^{-9}$. We estimate $\kappa_{10}$ using the dust properties for uncoagulated silicate dust as tabulated by \cite{Ossenkopf1992}\footnote{https://hera.ph1.uni-koeln.de/$\sim$ossk/Jena/tables/mrn0}, which provide the values of $\kappa(\lambda)$ for a standard MRN distribution. Since the grains in our model have a radius $a=0.01$\,\micron, which is $\sim20$\% larger than the density-weighted grain radius for MRN ($a=0.0083$\,\micron), and $\kappa \propto a^{-1}$, our adopted value of $\kappa_{10}=1.72\times10^3$\,cm$^{2}$g$^{-1}$, is obtained by scaling the tabulated value of $\kappa_{10}=2.07\times10^3$\,cm$^{2}$g$^{-1}$ by 0.0083/0.01. We derive a total mass of amorphous silicate dust of $1.86\times10^{-6}$\,\me.

\begin{figure}

\includegraphics[width=0.8\textwidth]{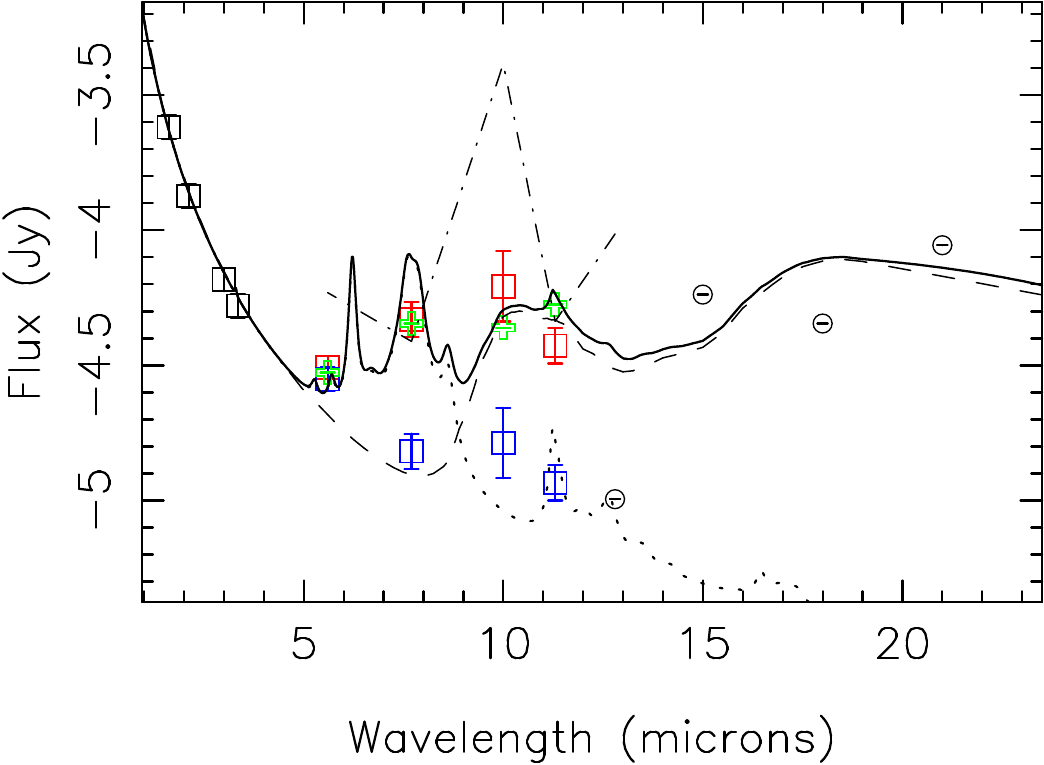}
\vskip -0.01in

\caption{Observed mid-IR photometry (black symbols) and model SED (smooth curve) for the CSPN of NGC\,6720 and a circumstellar dust shell with small silicate grains and PAHs. For $\lambda<5$\,\micron, black boxes show the total flux. For $\lambda>5$\micron, red (blue) boxes show the total (core) flux. Dotted curve shows the SED of the PAH emission, and the dashed curve shows the SED of the small silicate grains (+ the central star). Circles show upper limits for the core flux. Error bars on the observed photometry are conservative estimates. Dash-dot curve shows the relative fluxes of a patch covering a region of nebular emission in the near-vicinity of the CS. Green square symbols show the band-averaged model fluxes.}

\label{cloudy-mod-sed}
\end{figure}

\begin{figure}

\includegraphics[width=0.8\textwidth]{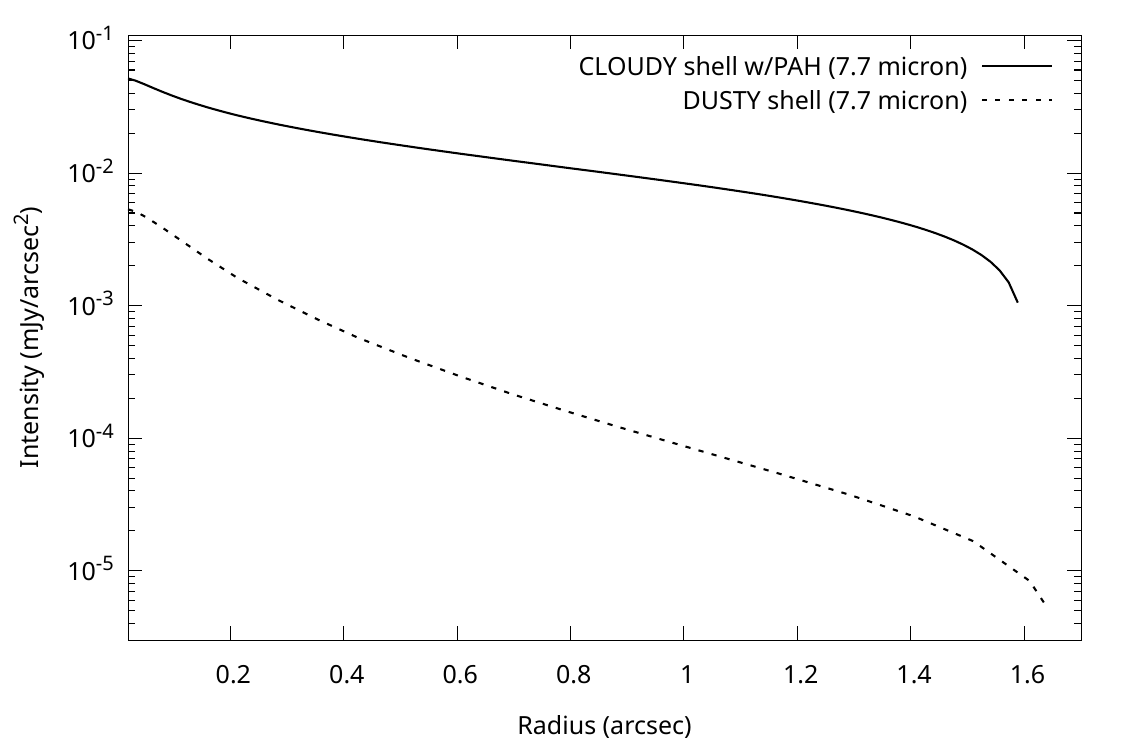}
\caption{The $7.7$\,\micron~radial intensity from the best-fit CLOUDY model (small silicate grains and PAHs) compared to that from the best-fit DUSTY model  (small silicate grains only).}
\label{cloudy-mod-770w}
\end{figure}

\subsection{Stochastic Emission: CLOUDY}\label{pah-cloudy}
We have used version C23.01 of the CLOUDY code \citep{ferland2017} to investigate whether the presence of PAHs in the dust cloud can help resolve the discrepancy between the data and our models at 7.7\,\micron. Using a dust shell with the same inner and outer radii, and density law, as derived from the DUSTY model, but with a dust composition consisting of charged PAH clusters with 120 atoms\footnote{using opacity file ph3c\_c120.opc in the CLOUDY C23.01 package}, we derive the SED of the emission from these grains (Fig.\,\ref{cloudy-mod-sed}, dotted curve) and add it to the SED derived from the DUSTY model (Fig.\,\ref{cloudy-mod-sed}, dashed curve). We find a very good fit to the observed 7.7\,\micron~flux (Fig.\,\ref{cloudy-mod-sed}, solid curve). In addition, the relatively smaller discrepancy that we found between the observed and DUSTY model 5.6\,\micron~flux, is now significantly reduced. The CLOUDY model $7.7$\,\micron~radial intensity is significantly higher than that in the DUSTY model at all radii, as expected, due to the contribution of the non-stochastic PAH emission (Fig.\,\ref{cloudy-mod-770w}). We do not attempt to further fine-tune the PAH model because a very large variety of PAH particles (e.g., with different numbers of C atoms) are likely to be present. We discuss the origin of PAHs in \S\,\ref{discuss} below. The total mass of PAH grains is $7.27\times10^{-7}$\,\me.

Inclusion of a small amount of gas, e.g., resulting from a tenuous hot stellar wind from the CSPN, with, for example, a density of (say) $\lesssim0.05$\,\vdensunit~at the inner radius of the dust shell (see below) produces negligible gaseous line emission, with no significant effect on the model photometry in the broad-band filters. The only gaseous line visible is that due to [Ne\,VI]\,7.65\,\micron; its integrated flux is very small compared to that of the much broader 7.7\,\micron~PAH feature. The model 3.3\,\micron~PAH feature is very weak and contributes negligibly to the flux in the F335M filter. We note that \cite{Wesson2024} find evidence for possible weak PAH emission in the F335M and F1000W filters in a narrow ring in the outer parts of the nebular shell, contributing $<14$\,\% and $<7$\,\% to the flux seen in these filters.

\subsection{Gas Emission: CLOUDY}\label{gas-cloudy}
The WD central stars of PNe are known to produce hot, line-driven winds that appear very early after the star leaves the AGB (e.g., for $T_{eff}\lesssim10$\,kK) and fade away on the WD cooling track (i.e., for $T_{eff}\gsim105$\,kK) (see Fig. 1 of \cite{krticka20}). The wind mass-loss rate depends mostly on the stellar luminosity (e.g., \citealt{Castor75,Sander17}). Models of such winds by \cite{krticka20} for a CSPN of mass 0.569\,\ms~produce a ``knee" in the cooling curve at $T_{eff}\sim117.8$\,kK, where the mass-loss rate has dropped to $2.6\times10^{-11}$\,\my, the wind speed is 1830\,\kms, and $L_{*}=10^3$\,\ls.  Subsequently, the WD enters a rapidly cooling phase, and the mass-loss rates drop very rapidly. For example, the CSPN mass-loss rate is $5.1\times10^{-13}$\,\my~for $T_{eff}\sim105.7$\,kK, and there is no wind as the WD cools further. In the case of NGC\,6720, the current (progenitor) mass is inferred to be $0.58\,(1.5-2)$\,\ms. The CSPN is believed to have reached a peak temperature (i.e., at the knee of the cooling curve) at $L_{*}=3\times10^3$\,\ls~(\cite{Wesson2024}) -- thus, given its current luminosity, it is now in the rapidly cooling phase (\cite{Wesson2024}). Hence, it appears likely then that the wind mass-loss rate for NGC\,6720's CSPN is $\lesssim10^{-13}$\,\my. The gas density due to such a wind (assuming a typical expansion velocity of 2000\,\kms) at the inner radius of the dust shell, 11.7\,au, is $\lesssim0.05$\,\vdensunit; the resulting line emission is insignificant compared to the dust emission.

\subsection{Possible Unresolved Companion}\label{lowmasscomp}
\begin{figure}[hbtp]
\vskip 0.01in
\includegraphics[width=0.8\textwidth]{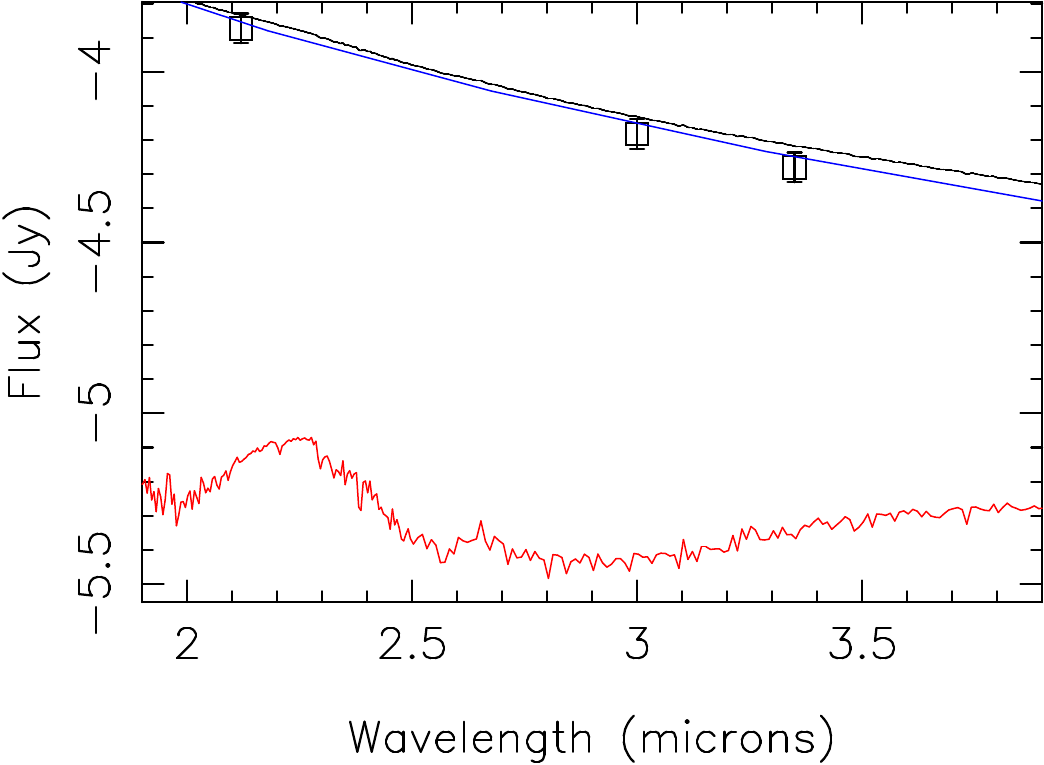}
\vskip -0.01in
\caption{Model SED of the CSPN (blue curve), an M8 MS companion (red curve), and the total SED (CSPN + M8) (black curve), together with the observed photometry (black rectangles).}
\label{comp_sed}
\end{figure}
We can constrain the mass of any unseen (unresolved) stellar companion, e.g., such as CSPN(C), from the SED of the CS over the wavelength range where dust emission does not contribute significantly (i.e., $<5$\,\micron). We have investigated the effect of including the theoretical spectrum of such a companion with a range of spectral types M7.5V or later, since for an M7.5V companion, the resulting flux at 3.35\,\micron~is 20\% more than observed, well above the observed upper limit. The values of luminosity and effective temperature for late-type MS dwarfs have been taken from \cite{Cifuentes2020}. The model spectra have been extracted from the BT-NextGen (AGSS2009) dataset (\citealt{Allard2011}) archived at http://svo2.cab.inta-csic.es/theory/newov2/index.php. The results of this modelling (Table\,\ref{tbl-ms-comp}) show that the highest-mass unresolved MS companion that can be present and remain undetected is less massive than a M-dwarf with spectral type M9.5V, with L$\sim2.35\times10^{-4}$\,\ls~and $T_{eff}\sim2300$\,K. An M9.5V companion increases the model SED flux in the F356M filter by 13\%, above the upper limit on the observed F335M flux. Noting that the model luminosities are also uncertain, we find that if we peg the M7.5V and M8.0V models at their lower-limit luminosities, the resulting F356M model flux values are still 18\% and 15\% above the observed values. We conservatively conclude that a companion, if present, is of spectral type M8.0 or later, implying a mass $\le0.1$\,\ms. Fig.\,\ref{comp_sed} shows model SEDs of an M8 MS companion and the total SED (CSPN + M8), together with the observed photometry in the wavelength range that is most sensitive to low-mass companions.

\section{Central Star Photometric Variability}\label{cs-var}
For the CS of NGC\,6720, the value of the Gaia variability flag, ``phot\_variable\_flag", is VARIABLE. Light curves for it are available in the Gaia photometric bands G, Bp and Rp\footnote{Gaia DR3 Part 1. Main source: I/355 -- https://cdsarc.cds.unistra.fr/viz-bin/cat/I/355 \citep{Gaia2022}}.
We have downloaded and analyzed these light curves, which cover a period of $\sim$950 d. We first examined the G-band data, as these have the highest S/N. The cadence is such that there are multiple pairs of successive datapoints that are very close in time ($<0.2$ d) (``close-time-clustered" datapoints), compared to the median time interval ($\gsim$30 d). In addition there was one point for the which the fractional flux error was much larger than the median fractional flux error, i.e., by 6.5$\sigma_{err}$, where $\sigma_{err}$ is the standard deviation of the fractional flux errors. We rejected this point, reducing $\sigma_{err}$ by a factor 2.6 in the resulting dataset (with 49 datapoints in total). We then removed two outliers with flux values outside $\pm$3$\sigma$ of the error-weighted mean flux, $F(G)_{ave}=10397$ e$^{-}$/s, further reducing the error-weighted standard deviation in the flux, $\sigma_{F(G)}$ from 108 to 85 (Fig.\,\ref{cs-lc-aver}, top, cyan symbols). The fluxes of the close-time-clustered datapoints (Fig.\,\ref{cs-lc-aver}, bottom) were averaged to produce a final dataset of 28 datapoints (with $\sigma_{F(G)}=82$) (Fig.\,\ref{cs-lc-aver}, top, red symbols). The statistics of the various datasets are summarised in Table\,\ref{tbl-gband} -- the value of $F(G)_{ave}$ does not change significantly across these datasets.

\begin{figure}[hbtp]
\vskip -2.0in
\includegraphics[width=0.9\textwidth]{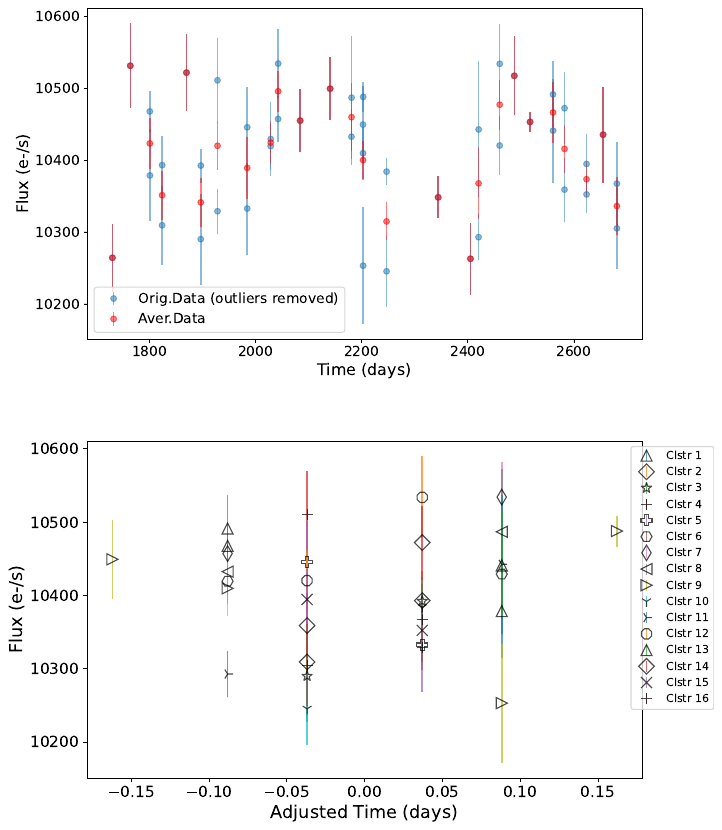}
\vskip -2.1in
\caption{(top) Gaia DR3 G-band count rate as a function of time for the CS of NGC\,6720: original data with one bad datapoint and two 3$\sigma$ outliers removed (cyan symbols), and with close-time-clustered points averaged (red symbols); (bottom) the G-band count rates for the close-time-clustered data points, with the mean time for each cluster set to 0.}
\label{cs-lc-aver}
\end{figure}

The flux variations of the CS are relatively small -- for the outliers-removed dataset, we find $\sigma_{F(G)} / F(G)_{ave} = 0.0082$.
We therefore first consider whether scan-angle-dependency of Gaia epoch photometry (\citealt{Holl2023}) could be responsible for producing spurious variability in the G band. \cite{Holl2023} provide two important parameters for assessing whether the G-band photometric variability resulting from scan-angle-dependency, labelled spearmanCorrIPDgFoV (also $r_{ipd}$) and spearmanCorrExfgFoV (also $r_{exf}$) in their variability catalog (https://doi.org/10.26093/cds/vizier.36740025: \citealt{Hollvizier23}). Relatively high absolute values of these parameters\footnote{the range is $0-1$}, i.e., |$r_{ipd}$| and |$r_{exf}$| indicate that the variability is spurious, and several studies that have extracted astrophysically variable sources using the G-band photometry (e.g., \citealt{Lebzelter2023, Mowlavi2023, Carnerero2023, Distefano2023}) have rejected sources for which |$r_{ipd}$|$>0.7$ and |$r_{exf}$|$>0.7$. However, since the values of these parameters for the CS of NGC\,6720 are very small ($r_{ipd}=0.12$ and $r_{exf}=-0.16$), we conclude that its observed flux variabilty is real.

The final dataset was subjected to Lomb-Scargle periodogram analysis, revealing a strong peak correponding to a period of $P=383$\,d, with an amplitude and phase of, respectively, 66.4 e$^{-}$/s (0.64\% of the median flux) and 0.20, and a false-alarm probability, FAP=0.2\footnote{the probability of observing the peak in a random, noise-only time series (i.e., a lower FAP value indicates that the peak is more significant)}. We then made a sinusodial fit of the model to the data, and then a few datapoints that were found to be offset by more than $3\sigma$ from the fitted curve  (a total of 3), were removed from the data set, and a second Lomb-Scargle periodogram analysis was carried out. This resulted in a slightly stronger peak at $P=371$\,d (Fig.\,\ref{cs-lc-g}), with an amplitude and phase, respectively, of 59.7 e$^{-}$/s (0.57\% of the median flux) and the same FAP.

A FAP value $<0.05$ is generally considered statistically significant, whereas a FAP value $>0.1$ indicates a less significant peak, likely due to noise. The FAP value is calculated based on the assumption that the time series is noise-only; if there are other sources of variability in the data, the FAP may not be accurate.

The light curves in the Bp and Rp filters (Figs.\,\ref{cs-lc-bp-rp}) are of lower signal-to-noise than in the G-band filter, and do not show an obvious periodicity.

In order to investigate whether the possible periodicity that we observe in the G-band filter lightcurve is an instrumental artifact, we have inspected the light curves of 6 field stars within a 5 arcmin radius of the CS that have similar G-band magnitudes or are slightly brighter (Table\,\ref{tbl-fs-lc}). The field stars show similar or lower variability amplitudes (Fig.\,\ref{fs-lc}); these flux variations are real because the values of $r_{ipd}$ and $r_{exf}$ for these sources are relatively small. We have carried out Lomb-Scargle periodogram analysis on these and we find that none of them show evidence for significant periodicity. We conclude that the G-band periodicity that we find for the CS of NGC\,6720, if significant, is most likely of astrophysical origin, and not an artifact.

The very short-term time variations of the G-band flux, as shown by the scatter of the fluxes in close-time-clustered points, are comparable to the longer-term varations, suggesting that, either the CS (i) has only short-term variations, and the (low-signficance) 383\,d period is not real, or (ii) has both a short-term variation and a longer term variation that is periodic, resulting in a relatively high FAP for the latter in the Lomb-Scargle analysis.

The mass of any close companion that could be responsible for the variability must be  $\le0.1$\,\ms, based on our modeling of the SED (\S\,\ref{lowmasscomp}).
Assuming that the 371\,d periodicity is real and due to orbital motion of he companion, the latter would (on the average) be separated by 0.9\,au from the CSPN.

\begin{figure}[hbtp]
\vskip -0.01in

\includegraphics[width=0.7\textwidth]{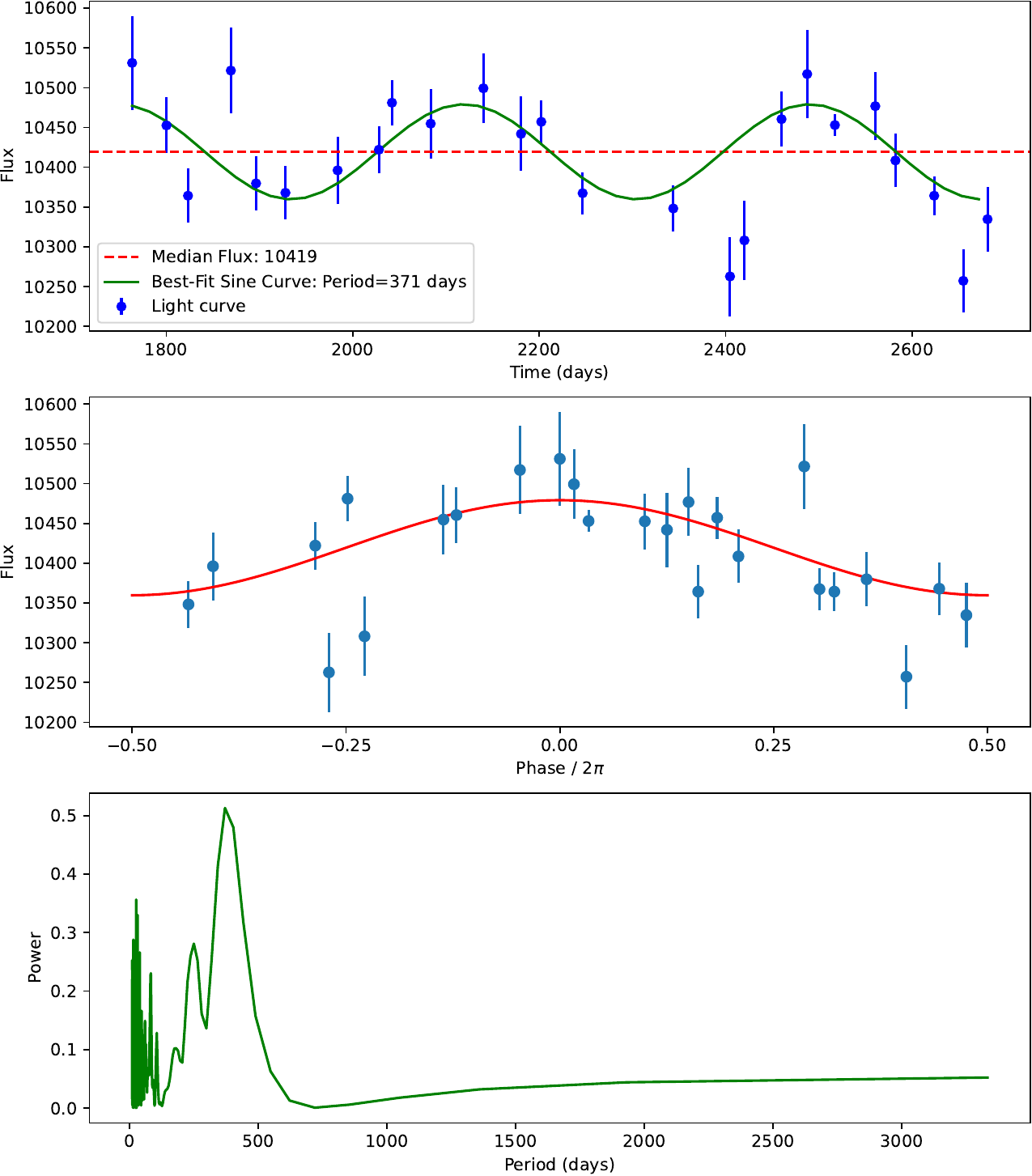}
\vskip -0.01in
\caption{(top) Gaia DR3 G-band light curve of the CS of NGC\,6720, overlaid with the best-fitting sinusoidal model, with period, P=371\,d; (middle)  Phase-folded Gaia DR3 G-band light curve of the CS of NGC\,6720, overlaid with the best-fitting sinusoidal model; (bottom) Lomb-Scargle periodogram of the light curve in the top panel.}
\label{cs-lc-g}
\end{figure}

\begin{figure}[hbtp]
\includegraphics[width=0.4\textwidth]{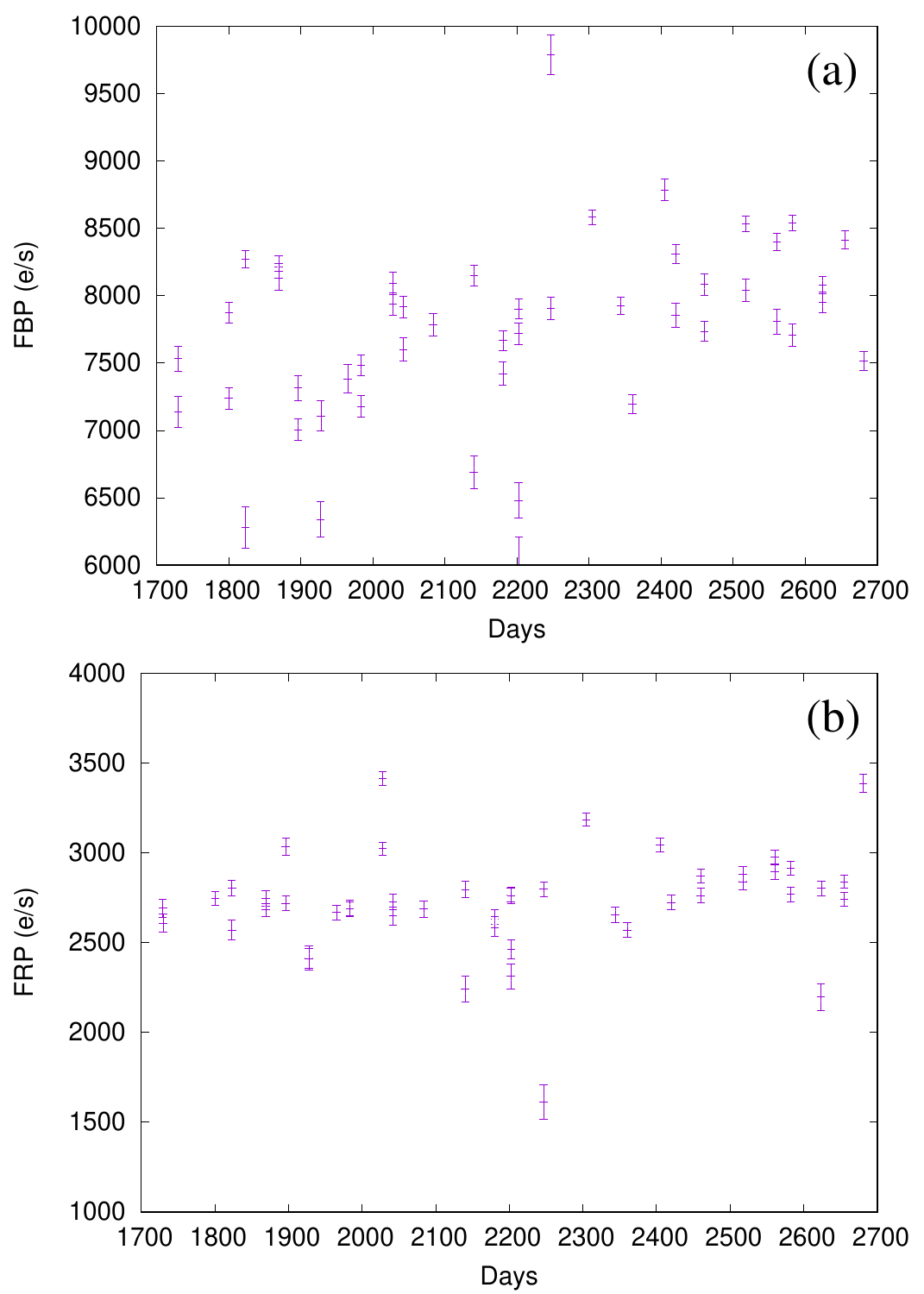}
\vskip 0.01in
\caption{Light curves of the CS of NGC\,6720 in the (a) Gaia DR3 BP-filter, (b)  Gaia DR3 RP-filter.}
\label{cs-lc-bp-rp}
\end{figure}

\begin{figure}[hbtp]
\vskip 0.1in
\includegraphics[width=0.7\textwidth]{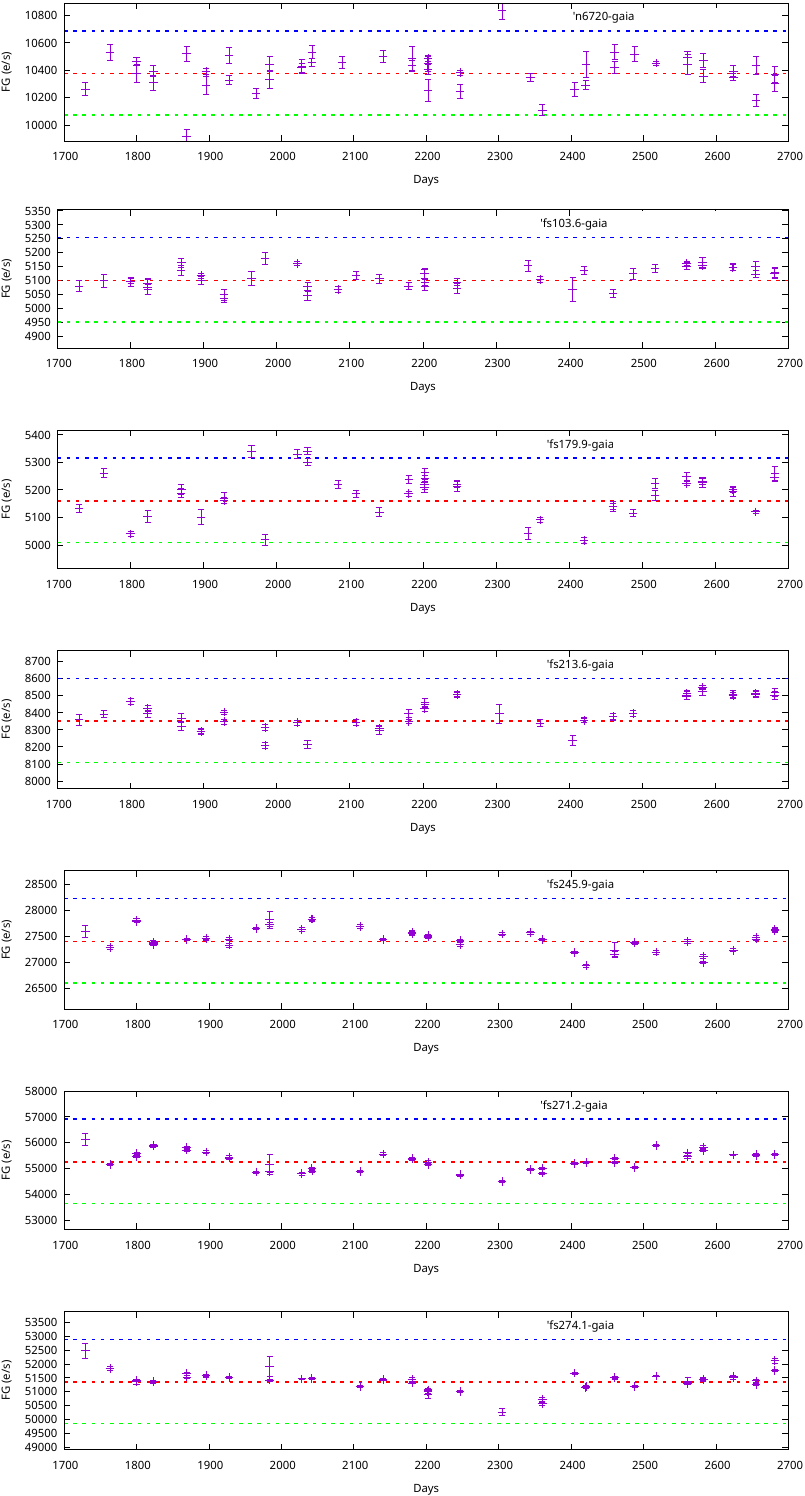}
\vskip -0.05in
\caption{Gaia DR3 G-filter light curve of the CS of NGC\,6720 compared with that of six field stars. In each panel, red dashed lines show the mean flux (e/s), whereas the blue (green) lines shows a flux value that is 3\% more (less) than the mean.}
\label{fs-lc}
\end{figure}

\section{Discussion}\label{discuss}
If the dust cloud around the CSPN includes the presence of PAHs, as we have proposed above, these will require continuous replenishment. This is because the strong UV radiation field of the CSPN is likely to photodissociate PAHs (which are generally small, $\lesssim5-10$\,\AA) very quickly. Given a typical dissociation energy of a bond within a PAH of $\sim5$ eV, and a photon rate of $\sim2\times10^{46}$ ph\,s$^{-1}$, a 50-carbon PAH, taking an FUV absorption cross section of $3.5\times10^{-16}$\,cm$^2$ (\citealt{Tielens08}), will be fully destroyed in about 1.5 hr at a typical radius of 600 AU, in the extended part of the dust cloud around the CSPN.

The replenishment can occur via the outgassing of cometary bodies and/or the collisional grinding of planetesimals (\citealt{Seok15,Seok17}). In addition to outgassing, UV irradiation can generate dust particles that are larger than PAHs, but which can also emit the broad features generally labelled aromatic infrared bands (AIBs) (\citealt{Kwok2001,Kwok2022}) -- these include a very strong, broad feature at $\sim$8\,\micron. Such dust grains consist of mixed aromatic/aliphatic organic nanoparticles (MAONs), a carbonaceous compound containing aromatic rings of different sizes and aliphatic chains of different lengths and orientations arranged in a 3-D amorphous structure. \citep{Kwok2001} suggest that UV irradiation continuously transforms aliphatic to aromatic groups in MAONs.

Such outgassing and/or collisional grinding of planetesimals, left-over from the MS phase of the primary, has been proposed as one mechanism to explain the presence of the compact dust cloud around the CSPN of the PN NGC\,3132, resulting from the dynamical evolution of a triple-star system (\citealt{Sahai2023}). As in the case of NGC\,3132, it is possible that the three stars, CSPN(A), CSPN(B), and CSPN(C) formed a stable hierarchical triple system while all the stars were on the MS (with an inner binary and a more distant tertiary), but became dynamically active on much longer time-scales due to the  ``Eccentric Kozai-Lidov" (EKL) mechanism (\citealt{Kozai1962,Lidov1962,Naoz2016}). It has been shown that the EKL mechanism can cause the inner binary to undergo large-amplitude eccentricity and inclination oscillations, driving it to have very small pericenter distances and even to merge (e.g., \citealt{Prodan2015,Stephan2018}), and the tertiary to move out to a large orbit or become unbound.

Alternatively, the dust cloud in the CS of NGC\,6720 could result from a strong binary interaction, when the primary was an RGB or AGB star, leading to the formation of stable circumbinary disks of gas and dust in Keplerian rotation (e.g., \citealt{vanwinckel2018, kluska2022}) with total masses (gas + dust) in the range $few\times10^{-3}$ to $10^{-2}$\,\ms~(e.g., \citealt{gallardocava2021}).
In contrast, the total dust mass that we derive for NGC\,6720's CS is relatively low ($\sim2.6\times10^{-6}$\,\me). It is much less than even the masses of the large asteroids in the Solar System, e.g., Ceres ($1.57\times10^{-4}$\,\me), Vesta ($4.34\times10^{-5}$\,\me) and Pallas ($4.34\times10^{-5}$\,\me), larger than those of small asteroids such as Siwa ($2.5\times10^{-7}$\,\me), and much larger than carbonaceous asteroids such as Bennu ($1.31\times10^{-14}$\,\me). In summary, if the dust cloud in NGC\,6720's CS is indeed a remnant of the disk resulting from binary interaction during an earlier evolutionary phase of the CSPN, it is clear that such a disk has now been almost completely dissipated.

The presence of a very close companion to CSPN(A) is suggested by the significant photometric variability of the CS (irrespective of whether or not a periodic signal is present)  (e.g., \citealt{Gladkowski24, Ali23a, Aller20}). Such ``extrinsic" variability may arise due to the operation of different mechanisms such as irradiation of a cold MS companion by the hot CSPN, ellipsoidal variability, and eclipses. Assuming that the orbital plane of the companion is the same as the equatorial plane of NGC\,6720, and the nebula is viewed nearly pole-on, eclipses of the CS by the companion are an unlikely explanation. This very close companion could either be CSPN(C), having moved inwards as a result of the EKL mechanism, or it could be a 3rd companion of CSPN(A). However, the photometric variability of the CS could also be intrinsic to the CSPN due to non-radial pulsations or spots on the surface of the WD. For example, a recent survey of Gaia DR3, TESS, and ZTF data reveal flux variations with periodicities from minutes to days (\citealt{Steen24}) in a sample of 105 WDs that have been attributed to these phenomena.

\section{conclusions}\label{conclude}
We have used JWST imaging in the near- to mid-IR wavelength range to investigate the central star of the PN NGC\,6720 and its close vicinity. Our main findings are as follows:
\begin{enumerate}
\item The central star is surrounded by a compact dust cloud -- evidence for this cloud comes from excess emission seen in the SED at wavelengths $\gsim$5\,\micron~as well as radially-extended emission in the 7.7, 10 and 11.3\,\micron~images.
\item We have modelled the SED spanning the UV to mid-IR wavelength range using the DUSTY radiative transfer code. The UV to near-IR wavelength SED shows the presence of a CSPN of luminosity 310\,\ls, with a line-of-sight interstellar extinction of $A_V=0.15^{\rm mag}$, assuming a stellar effective temperature of 135\,kK based on published studies. Our best-fit model provides a good fit to the radial intensity distributions of the CS at 10.3 and 11\,\micron, and shows that the dust cloud has a size of $\sim$2600\,au and consists of relatively small amorphous silicate dust grains (radius $\sim$0.01\,\micron) with a total mass of $1.9\times10^{-6}$\,\me. However, this model shows a very significant lack of extended emission at 7.7\,\micron.
\item We find, using the CLOUDY radiative transfer code, that in order to fit the 7.7\,\micron~emission, we require a smaller (but uncertain) mass, $7.3\times10^{-7}$\,\me, of (stochastically-heated) ionized PAHs, excited by the UV radiation from the CSPN. Since the same radiation also rapidly destroys PAH molecules, we speculate that these are likely being continuously replenished via the outgassing of cometary bodies and/or the collisional grinding of planetesimals.
\item We find significant photometric variability of the CS that could be due to the presence of a close MS dwarf companion of mass $\le$0.1\,\ms.
\end{enumerate}

\noindent{{\it Acknowledgements}: }
This work is based on observations made with the NASA/ESA/CSA James Webb Space Telescope, as part of GO program ID\,1558. The data were obtained from the Mikulski Archive for Space Telescopes at the Space Telescope Science Institute, which is operated by the Association of Universities for Research in Astronomy, Inc., under NASA contract NAS 5-03127 for JWST.\\

RS’s contribution to the research described here was carried out at the Jet Propulsion Laboratory, California Institute of Technology, under a contract with NASA, and was funded in part by NASA/STScI award JWST-GO-01558.002-A.  M.B. acknowledges support from European Research Council(ERC) Advanced Grant number SNDUST 694520. J.C. and E.P. acknowledge support from the University of Western Ontario, the Institute for Earth and Space Exploration, the Canadian Space Agency (CSA, 22JWGO1-14), and the Natural Sciences and Engineering Research Council of Canada. N.L.J.C. and J.B.-S. contributed to this work in the framework of the Agence Nationale de la Recherche's LabCom INCLASS (ANR-19-LCV2-0009), a joint laboratory  between ACRI-ST and the Institut d'Astrophysique Spatiale (IAS).  H.L.D. acknowledge support from grant JWST-GO-01558.003 and NSF grants AAG-175332 and AAG-2307117. M.M. and R.W. acknowledge support from the STFC Consolidated grant (ST/W000830/1). K.J. acknowledges support from the Swedish National Space Agency. PJK acknowledges financial support from the Research Ireland Pathway programme under Grant Number 21/PATH-S/9360. A.M. acknowledges the support from the State Research Agency (AEI) of the Ministry of Science, Innovation and Universities (MICIU) of the Government of Spain, and the European Regional Development Fund (ERDF), under grants PID2020-115758GB-I00/AEI/10.13039/501100011033 and PID2023-147325NB-I00/AEI/10.13039/501100011033; his contribution is based upon work from COST Action CA21126 - Carbon molecular nanostructures in space (NanoSpace), supported by COST (European Cooperation in Science and Technology). AAZ acknowledges funding from the European Union OSCARS program under Grant Agreement no. 101129751, project 01-358.\\

\noindent{{\it Data Availability}: }
JWST raw and pipeline-calibrated data is available from MAST (program ID\,1558, \dataset[doi:10.17909/bv01-qg73]{https://doi.org/10.17909/bv01-qg73}).

\clearpage
\vskip 0.3in

\setlength{\bibsep}{3pt plus 0.3ex}

\begin{table*}

\caption{Photometry of the CS of NGC\,6720}
\label{tbl-flux-cs}
\begin{tabular}{llcccccc}
\hline     
Filter & Wavelength & Flux   & Error\tablenotemark{a} & Apert.\tablenotemark{b}     & Apert.\tablenotemark{c} & Apert.\tablenotemark{d} & Phot\tablenotemark{e}  \\
       & (\micron)  & (mJy)  & (\%)                   & Rad.($''$)                  & Corr.                   & PA ($\arcdeg$)          & Ref. \\
\hline
B      & 0.43  & 2.80   & 10 &... & ...  & ... & 15.405\tablenotemark{f}   (2)\\ 
V      & 0.555 & 1.79   & 10 &... & ...  & ... & 15.769\tablenotemark{f}   (2)\\ 
G-band & 0.622 & 1.78  &  10 & ... & ...  & ... & 15.646                   (4) \\  
R      & 0.71  & 1.34   & 10 &... & ...  & ... & 15.901\tablenotemark{g}   (1)\\ 
I      & 0.798 & 0.92   & 10 &... & ...  & ... & 16.602\tablenotemark{g}   (1)\\ 
J      & 1.235 & 0.45   & 10 &... & ...  & ... & 16.40\tablenotemark{g}    (1)\\ 
F162M  & 1.62  & 0.241  &  10 & $0\farcs50$ & 1.0  &  ... & (5)\\
F212N  & 2.12  & 0.134  &  10 & $0\farcs30$ & 0.94 &  ... & (5)\\
F300M  & 3.0   & 0.066  &  10 & $0\farcs50$ & 0.95 &  ... & (5)\\
F335M  & 3.35  & 0.0525 &  10 & $0\farcs50$ & 0.94 &  ... & (5)\\

F560W  & 5.60  & 0.028\tablenotemark{h} &  10 & $0\farcs14$ & 0.31 &  203--323 & (5)\\
F560W  & 5.60  & 0.031\tablenotemark{j} &  10 & $1\farcs8$ & ... &  203--323 & (5)\\

F770W  & 7.70  & 0.015\tablenotemark{h} &  15 & $0\farcs18$ & 0.50 &  178--268 & (5)\\
F770W  & 7.70  & 0.047\tablenotemark{j} &  15 & $1\farcs8$ & ... &  178--268 & (5)\\

F1000W & 10.0  & 0.016\tablenotemark{h} &  30 & $0\farcs23$ & 0.50 &  156--180 & (5)\\
F1000W & 10.0  & 0.058\tablenotemark{j} &  30 & $1\farcs8$ & ... &  156--180 & (5)\\

F1130W & 10.0  & 0.012\tablenotemark{h} &  30 & $0\farcs24$ & 0.47 &  156--180 & (5)\\
F1130W & 10.0  & 0.037\tablenotemark{j} &  30 & $1\farcs8$ & ... &  156--180 & (5)\\
F1280W & 11.3 & $<$0.010\tablenotemark{k} & ... & $0\farcs23$  & ... & ... & (5)\\
F1500W & 15.0 & $<$0.058\tablenotemark{k} & ... & $0\farcs28$  & ... & ... & (5)\\
F1800W & 18.0 & $<$0.045\tablenotemark{k} & ... & $0\farcs34$  & ...  & ... & (5)\\
F2100W & 21.0 & $<$0.088\tablenotemark{k} & ... & $0\farcs39$  & ...  & ... & (5)\\

\hline
\end{tabular}\\

\tablenotetext{a}{Percentage Error in Flux in Col.\,(3)}
\tablenotetext{b}{Radius for aperture photometry, or outer radius used for integrating radial intensity to determine flux in filters where CS emission is extended -- for the latter, no aperture correction is required}
\tablenotetext{c}{Aperture correction (for aperture photometry) determined using field stars}
\tablenotetext{d}{Position angles defining the full angular wedge used to extract radial intensity}
\tablenotetext{e}{References for photometry: (1) de Marco et al. (2013), (2) Hubble Source Catalog V.3 (Whitmore et al. 2016), (4) Gaia DR3 \citep{Gaia2022}, (5) this work; when photometry reference provides magnitudes, these are listed here}
\tablenotetext{f}{AB Magnitude}
\tablenotetext{g}{Vega Magnitude}
\tablenotetext{h}{Core Flux, $F_{core}$, derived from integration of radial intensity to radius $0.5\times$FWHM of PSF}
\tablenotetext{j}{Flux derived from integration of radial intensity to outer radius in Col.\,(5)}
\tablenotetext{k}{Upper limit for core flux derived using method described in \S\ref{making-sed}}
\end{table*}

\begin{table*}

\caption{Field Stars used for generating PSFs and Aperture Corrections}
\label{tbl-fs-psf}
\begin{tabular}{lcccc}
\hline
Star & Filter & RA(J2000.0)  & Dec(J2000.0)   \\

\hline
fs1 &  F560W   & 18:53:30.782  &  +33:01:42.70 \\
fs3 &  F770W   & 18:53:39.842  &  +33:01:46.30 \\
fs4 &  F1000W  & 18:53:40.775  &  +33:01:40.90  \\
fs1 &  F1130W  & 18:53:30.782  & +33:01:42.70 \\
\hline
\end{tabular}\\
\end{table*}

\begin{table*}
\caption{Photometry of Circular Patch near the CS of NGC\,6720}
\label{tbl-flux-patch}
\begin{tabular}{ll}
\hline     
Filter & Flux   \\
       & (mJy)  \\
\hline
F560W  & 0.0039 \\
F770W  & 0.0026 \\
F1000W & 0.028 \\
F1130W & 0.003 \\
\hline
\end{tabular}\\
\end{table*}

\begin{table}
\caption{Best-Fit Models of the Dust Emission towards the CS of NGC\,6720}
\label{tbl-dusty}

\begin{tabular}{lllllllll}
\hline
T$_{\rm d}$(in)\tablenotemark{a} & $R_{in}$\tablenotemark{b} & T$_{\rm d}$(out) &$R_{out}$\tablenotemark{c} &  n\tablenotemark{d}  & $\tau _V$\tablenotemark{e} &  $F_{\rm bol}$  & Dust.Comp.& M$_{\rm d}$\tablenotemark{f} \\
(K)                              & (arcsec, au)              & (K)              &(arcsec, au)               &                      &                            & (\fbolunit)     &           & (M${_\oplus}$) \\
\hline
1500                            & $0\farcs013$, 10.5         & 151              & $1\farcs66$, 1310        & $-0.8$                & $1.3\times10^{-8}$         & $1.6\times10^{-6}$ & silicate  & $1.86\times10^{-6}$ \\
1688                            & ...                        & 238              & ...                      & ...                   & ...                        & ...                & PAH       & $7.27\times10^{-7}$ \\
\hline
\end{tabular}
\tablenotetext{a}{The (input/output) dust temperature at shell inner radius for silicate/PAH dust}
\tablenotetext{b}{The (output) dust temperature at shell outer radius}
\tablenotetext{b}{The (inferred) inner radius of the dust shell}
\tablenotetext{c}{The (input) outer radius of the dust shell}
\tablenotetext{d}{The (input) exponent of the density power law ($\rho_d(r)\propto r^{-n}$) in the dust shell}
\tablenotetext{e}{The (input) dust shell's optical depth at $0.55\micron$}
\tablenotetext{f}{The (inferred) circumstellar dust mass}
\end{table}

\begin{table*}
\caption{Gaia G-band Time Series of the CS of NGC\,6720: Statistics}
\label{tbl-gband}
\begin{tabular}{lccc}
\hline     
Dataset               & No. of      &  Wtd.\,Mean & Std.\,Dev \\
Decr.                 & DataPts &  (e$^-$/s)  &  (e$^-$/s)  \\
\hline
Original              & 50 & ...   & ... \\
Bad-data removed      & 49 & 10397 & 107.6 \\
Outliers removed      & 47 & 10401 & 85.1  \\
CloseDataPts averaged & 28 & 10398 & 81.6  \\
\hline
\end{tabular}\\
\end{table*}

\begin{table*}
\caption{Field Stars with G-band light curves near the CS of NGC\,6720}
\label{tbl-fs-lc}
\begin{tabular}{llcccccc}
\hline     
Name & Offset   & RA(J2000)  & Dec(J2000)   & Gmag\tablenotemark{a}  & phot\_variable\_flag & $r_{ipd,G}$\tablenotemark{b} & $r_{exf,G}$\tablenotemark{c} \\
     & arcsec   & hh mm ss  & dd mm ss & mag \\
\hline

FS1 & 103.578 & 18 53 31.9920361623 & +33 03 21.015084474 & 16.416496 & VARIABLE & -0.030698 & -0.278665\\
FS2 & 179.898 & 18 53 45.9813883771 & +33 03 41.753314212 & 16.401575 & VARIABLE & -0.074560 & -0.189189\\
FS3 & 213.640 & 18 53 32.9190411134 & +32 58 13.093566670 & 15.875365 & VARIABLE & -0.173693 & 0.189916\\

FS4 & 245.875 & 18 53 21.6164058659 & +33 04 43.133991427 & 14.591276 & VARIABLE & 0.184214 & 0.112892\\
FS5 & 271.235 & 18 53 16.7110519429 & +33 04 07.034403143 & 13.829674 & VARIABLE & -0.056917 & -0.150474 \\
FS6 & 274.113 & 18 53 35.1056734346 & +32 57 11.036167760 & 13.910662 & VARIABLE & -0.045757 & -0.047561\\

\hline
\end{tabular}\\
\tablenotetext{a}{Gaia DR3 G-band mean magnitude}
\tablenotetext{b}{spearmanCorrIPDgFoV from table J/A+A/674/A25/vspursig}
\tablenotetext{c}{spearmanCorrExfgFoV from table J/A+A/674/A25/vspursig}
\end{table*}

\begin{table*}
\caption{Effect of a Main-Sequence Stellar Companion on the SED of the CS of NGC\,6720}
\label{tbl-ms-comp}
\begin{tabular}{lcccc}
\hline     
Spectral\tablenotemark{a} & T$_{eff}$\tablenotemark{a} & Luminosity\tablenotemark{a} & Mass\tablenotemark{a} & F$_{mod}$(3.35\,\micron)/F$_{obs}$(3.35\,\micron) \\
Type & (K)       & ($10^{-4}$\,\ls)  & ($\ms$)  & \\
\hline     
M7.5V   & 2500 $\pm$82   & 5.8 $\pm$1.2   & 0.104 $\pm$0.009  & 1.20 \\
M8.0V   & 2500 $\pm$91   & 5.1 $\pm$1.6   & 0.104 $\pm$0.014  & 1.19 \\
M9.5V   & 2300 $\pm$45   & 2.69 $\pm$0.35 & 0.077 $\pm$0.008 & 1.13 \\
L0.5V   & 2200 $\pm$61   & 2.17 $\pm$0.15 & 0.079 $\pm$0.004 & 1.13 \\
L1.5V   & 2000 $\pm$172  & 1.81 $\pm$0.35 & 0.094 $\pm$0.016 & 1.12 \\

\hline
\end{tabular}\\
\tablenotetext{a}{from \cite{Cifuentes2020}}
\end{table*}

\end{document}